\documentstyle[prb,aps,multicol,epsf]{revtex}
\widetext
\renewcommand{\narrowtext}{\begin{multicols}{2} 
\global\columnwidth20.5pc}
\renewcommand{\widetext}{\end{multicols} \global\columnwidth42.5pc}
\newcommand{\be}{\begin{eqnarray}}
\newcommand{\ee}{\end{eqnarray}}
\newcommand{\ba}{\begin{array}}
\newcommand{\ea}{\end{array}}
\newcommand{\no}{\nonumber}
\newcommand{\tr}{\mbox{tr}\,}
\newcommand{\trg}{\mbox{trg}\,}

\begin{document}
\draft
%
% comment out the following line for single column output
% \twocolumn[\hsize\textwidth\columnwidth\hsize\csname @twocolumnfalse\endcsname
%
\title{
Energy-level correlations in chiral symmetric disordered systems : \\
  Corrections to the universal results
}

\author{
Kazutaka Takahashi${}^{1,}$\footnote{present address: Theoretische Physik III, 
Ruhr-Universitaet Bochum, 44780 Bochum, Germany} and 
Shinji Iida${}^{2}$ 
}
\address{
${}^1$ Institute of Physics, University of Tsukuba,
 Tsukuba 305-8571, Japan\\
${}^2$ Faculty of Science and Technology, Ryukoku University,
 Otsu 520-2194, Japan\\
}

\date{\today}  
\maketitle
\begin{abstract}
 We investigate the deviation of the level-correlation functions
 from the universal form for the chiral symmetric classes.
 Using the supersymmetric nonlinear sigma model
 we formulate the perturbation theory.
 The large energy behavior is compared with 
 the result of the diagrammatic perturbation theory.
 We have the diffuson and cooperon contributions even in the average 
 density of states. 
 For the unitary and orthogonal classes 
 we get the small energy behavior that suggests
 a weakening of the level repulsion.
 For the symplectic case we get 
 a result with opposite tendency.
\end{abstract}
\narrowtext
%%%%%%%%%%%%%%%%%%%%%%%%%%%%%%%%%%%%%%%%%%%%%%%%%%%%%%%%%%%%%%%%%%%%%%%%
%%%%%%%%%%%%%%%%%%%%%%%%%%%%%%%%%%%%%%%%%%%%%%%%%%%%%%%%%%%%%%%%%%%%%%%%
\section{Introduction}
 Statistical properties of disordered systems have attracted interest
 since the work of Wigner and Dyson. 
 At present, we recognize that disordered systems 
 are classified with the notion of the symmetric space.\cite{z} 
 In addition to the three traditional classes of disordered 
 systems---unitary, orthogonal and symplectic---we have three classes for 
 chiral symmetric systems and four classes for 
 normal-superconducting systems.\cite{az} 
 
 The chiral symmetric Hamiltonian can be written as
\be
 H = \pmatrix{
 0 & \omega_1-i\omega_2 \cr
 \omega_1+i\omega_2 & 0 \cr},
\ee
 where $\omega_1$ and $\omega_2$ are random Hermitian matrices.
 This Hamiltonian is relevant for the motion of  
 a single electron in a lattice with random magnetic fields,\cite{rfm} 
 which has been under intensive study because of its relation 
 to the fractional quantum Hall effect close to half filling \cite{hall}
 and to the gauge theory of high $T_{c}$ superconductivity.\cite{tc}
 Further, the low energy properties of QCD are also believed 
 to be described by this type of Hamiltonian.\cite{sv}

 This Hamiltonian has pairs of eigenvalues with $\pm E$.
 The zero energy point becomes a special point 
 and new universality appears.\cite{sv,sn}
 This universal nature at the ergodic limit is studied very well
 and we know that even the average density of states 
 becomes universal.\cite{vz}
 
 In the ergodic limit, we are interested in phenomena 
 whose time scale is much larger than the diffusion time $1/E_c$ 
 where $E_c$ is the Thouless energy.
 Hence an electron has enough time to wander everywhere in the 
 system and the spatial structure of the system can be ignored. 
 If we want to discuss 
 relatively short-time phenomena (or long-range correlations of energies), 
 we need to leave the ergodic region and enter the diffusive region.
 
 In the diffusive region of the traditional three classes 
 of disordered systems, 
 it is well known that the diffuson and cooperon modes 
 that express the quantum interference effects play an important role. 
 Especially owing to these modes there appear the deviations from 
 the universal behavior.
 The aim of this paper is to give analogous results  
 for the chiral symmetric disordered systems.
 The deviations from the universal form 
 are interesting from the viewpoint of the 
 weak localization because we do not know whether or not the localization
 occurs for these classes. 
 Besides, mesoscopic devices with 
 random magnetic fields are now available.\cite{exp}
 
 In our previous work,\cite{th} we discussed these problems 
 in the context of QCD for the chiral unitary class
 but some of the result contain some errors. 
 In this paper, for all three -- chiral orthogonal, unitary and
 symplectic -- classes, we calculate the level 
 correlation functions.
 Using the Efetov's supersymmetry method,\cite{ef} 
 we formulate the perturbative expansion and calculate 
 the large energy asymptotics of the level-correlation functions.
 We derive the small energy behavior of the density of states 
 using the improved calculation developed by 
 Kravtsov and Mirlin.\cite{km}
 The large energy behavior is compared with 
 the results of the diagrammatic perturbation theory 
 to examine which contributions are important
 and which ones make differences between the chiral classes and 
 the traditional ones.
%%%%%%%%%%%%%%%%%%%%%%%%%%%%%%%%%%%%%%%%%%%%%%%%%%%%%%%%%%%%%%%%%%%%%%%%%
%%%%%%%%%%%%%%%%%%%%%%%%%%%%%%%%%%%%%%%%%%%%%%%%%%%%%%%%%%%%%%%%%%%%%%%%%
\section{Level correlation functions}
 We formulate the calculation of the level-correlation functions 
 by using the supersymmetry method.
 We consider the effective action 
\be
 S &=& \frac{\pi E}{2\Delta V}\int d^dx\,\trg Q(x)\Sigma_z 
 +\frac{\pi D}{4\Delta V}\int d^dx\,\trg [\nabla Q(x)]^2, \no\\
 \label{S}
\ee
 where trg denotes the supertrace\cite{ef} and $D$ is the diffusion constant 
 and $\Delta$ is the mean level spacing. 
 $V=L^d$ is the volume of the system and 
 $\Sigma_z=\mbox{diag}(1,1,-1,-1)$.
 We note that this is the effective action for the one-point function.
 The supermatrix $Q(x)$ that 
 parametrizes the saddle-point manifold 
 is a $4\times 4$ supermatrix for the chiral unitary class
 and satisfies $Q^2(x)=-1$.\cite{ast} 
 For the chiral orthogonal and symplectic classes, 
 the size of the matrix is duplicated. 
 The structure of the supermatrices comes from chiral symmetry 
 and supersymmetry (and the complex conjugation property).
 This nonlinear sigma model is derived 
 from the schematic model of random matrices.\cite{th} 
 A similar model is derived by using the replica method.\cite{gade} 
 Related models using the supersymmetry method 
 are derived for the random-flux model,\cite{rfm} 
 for a Dirac particle in gauge field disorder \cite{gww}
 and for a two-dimensional(2D) electron gas in a random magnetic field.\cite{te}
 Although Eq.(\ref{S}) is derived starting from a particular model, 
 we think this form is broadly valid 
 because we know, from the experience for the traditional classes, 
 the effective action does not depend on the details of specific models.
 In the case of $d=4$, a similar model has been discussed in great detail 
 in the context of QCD.\cite{chpt}
 The nonlinear sigma model was written down based on the symmetries
 of the theory and the nature of the diffusive regime as well as 
 the ergodic regime was studied.

 The average density of states can be calculated 
 from the following expression:
\be
 \left<\rho(E)\right> &=& -\frac{1}{4\Delta V}\,\mbox{Im}\int DQ
 \left[\int d^dx\,\trg kQ(x)\Sigma_z\right] \,\mbox{e}^{-S}, \no\\
\ee
 where $k=\mbox{diag}(1,-1)\otimes 1_2$.
 For the ergodic limit (which means we ignore the spatial dependence
 in the above expression), by parametrizing the saddle-point manifold,
 we can calculate the density of states
 and get the universal result.\cite{ast} 
 In this paper, for the formulation of the perturbative expansion,
 we take another parametrization as
\be
 Q(x) &=& -i\,\Sigma_z (1+P)(1-P)^{-1} \no\\
      &=& -i\,\Sigma_z 
 \left[\,1+2P+2P^2+2P^3+2P^4+\cdots\,\right],  \label{p}\\
 & & P = \pmatrix{
 0 & it(x) \cr -it(x) & 0 \cr},
\ee
 where $t(x)$ is a $2\times 2$ ($4\times 4$) supermatrix for 
 the chiral unitary (orthogonal and symplectic) class.
 This parametrization has the advantage that the measure is normalized.
 A similar parametrization is used 
 for the traditional classes.\cite{ef} 
 In previous works,\cite{th} we have taken another parametrization.
 The measure is not normalized and has been treated perturbatively.

 Using the parametrizaton (\ref{p}), 
 we can formulate the perturbation theory.
 The free action can be written as
\be
 S_0 &=& \frac{-i\pi E}{\Delta V}\int d^dx\,\trg P^2 
 -\frac{\pi D}{\Delta V}\int d^dx\,\trg (\nabla P)^2.
\ee
 The perturbative expansion is 
 expressed using the diffusion propagator
\be
 \Pi(E;x,y) &=& \sum_{k}\tilde{\Pi}(E,k)\,
 \mbox{e}^{ik(x-y)},\\
 \tilde{\Pi}(E,k) &=& \frac{1}{4\pi\left(
 \frac{D}{\Delta}\,k^2-i\,\frac{E}{\Delta}\right)}.
\ee
%%%%%%%%%%%%%%%%%%%%%%%%%%%%%%%%%%%%%%%%%%%%%%%%%%%%%%%%%%%%%%%%%
\subsection{Large-$E$ behavior of the density of states}
 The level-correlation function can be expressed 
 in powers of the diffusion propagator.
 This expansion is valid for the large $E$ (large $E/\Delta$).
 Therefore, we can see the large-$E$ asymptotics of the level-correlation
 function in this calculation.
 The calculation goes along the same lines as the traditional class.
 The only difference is the contraction rules.
 For the chiral unitary class, the contraction rules are expressed as
\be
 \left<\trg AP(x)BP(y)\right>_Q 
 &=& \frac{1}{2}\,\Pi(E;x,y)\,\Bigl[\,\trg A\,\trg B \no\\
 & & -\trg A\Sigma_z\,\trg B\Sigma_z \no\\
 & &  -\trg A\Sigma_x\trg B\Sigma_x \no\\
 & & +\trg A\Sigma_y\,\trg B\Sigma_y \,\Bigr], \label{cont1}\\
 \left<\trg AP(x)\,\trg BP(y)\right>_Q 
 &=& \frac{1}{2}\,\Pi(E;x,y)\,\trg
 \Bigl[\,AB \no\\
 & & - A\Sigma_zB\Sigma_z-A\Sigma_xB\Sigma_x \no\\ 
 & & +A\Sigma_yB\Sigma_y \,\Bigr], \label{cont2}
\ee
 where $A$ and $B$ are arbitrary $4\times 4$ supermatrices and 
 $\left<\cdots\right>_Q$ denotes the average 
 with respect to the free action $S_0$.
 $\Sigma_x$, $\Sigma_y$, and $\Sigma_z$ are $4\times 4$ Pauli matrices.

 We can see that the differences from the traditional unitary class
 \cite{ef} are the
 third and fourth terms of Eqs. (\ref{cont1}) and (\ref{cont2}).
 These differences come from the structure of the matrix $P$.
 For the traditional unitary class, the matrix $P$ can be written as
\be
 P = \pmatrix{
 0 & it_{12}(x) \cr -it_{21}(x) & 0 \cr}.
\ee
 The matrices $t_{12}$ and $t_{21}$ relate each other
 with $t_{21}=kt^{\dag}_{12}$.
 When we take the contraction, these matrices couple with each other
 but there are no self-couplings such as $\left<At_{12}Bt_{12}\right>$.
 Due to this difference, 
 we get the above results for the contraction rules.

 Using these rules, we obtain the density of states 
 for the chiral unitary class as
\be
 \left<\rho(E)\right> 
  &=& \frac{1}{\Delta}\,\mbox{Re}\biggl[\, 1 
 +2\,\Bigl\{\sum_k\tilde{\Pi}(E,k)\Bigr\}^2 \no\\ 
 & & +\frac{16\pi iE}{\Delta}\,
 \Bigl\{\sum_k\tilde{\Pi}(E,k)\Bigr\}\,
 \Bigl\{\sum_k\,\tilde{\Pi}^2(E,k)\Bigr\}
 +\cdots \,\biggr]. \no\\
 \label{1pt}
\ee
 For the ordinary class, this function just equals to $1/\Delta$.
 The nontrivial contribution appears due to the chiral structure.
 We note that the first-order correction ($\Pi^1$ term) vanishes.
 In the previous work,\cite{th} we have erroneously neglected 
 the higher-order terms coming from $(\nabla Q)^2$ and 
 have got a wrong result. 
 A similar term in the present parametrization gives no contribution.    

 If we take the $k=0$ mode only, we have
\be 
 \left<\rho(E)\right> 
 &=& \frac{1}{\Delta}\,\biggl[\, 1 
 +\frac{1}{8(\pi E/\Delta)^2}
 +\cdots \,\biggr],
\ee
 which coincides with an asymptotics of 
 the exact result at the ergodic limit.\cite{vz}

 Similar calculations can be performed 
 for the chiral orthogonal and symplectic classes.
 For the chiral orthogonal class, 
 the saddle-point manifold is parametrized by the $8\times 8$
 supermatrix $Q(x)$.
 We have additional condition
\be
 Q &=& C\,Q^T\,C^T ,
\ee
 where $C$ is the complex-conjugation matrix.
 Similar calculations apply for the chiral symplectic class 
 by the redefinition of the complex-conjugation matrix $C$.\cite{ef} 
 The contraction rules can be written as
\be
 \left<\trg AP(x)BP(y)\right>_Q &=& \frac{1}{4}\,\Pi(E;x,y)\,
 \Bigl[\,\trg A\,\trg B \no\\ 
 & & -\trg A\Sigma_z\,\trg B\Sigma_z \no\\
 & & -\trg A\Sigma_x\,\trg B\Sigma_x \no\\
 & & +\trg A\Sigma_y\,\trg B\Sigma_y \no\\
 & & +\alpha\ \trg \bigl(\,A\Sigma_xCB^TC^T\Sigma_x \no\\
 & & +A\Sigma_yCB^TC^T\Sigma_y \no\\
 & & +A\Sigma_zCB^TC^T\Sigma_z \no\\
 & & -ACB^TC^T \,\bigr)\,\Bigr], \\
 \left<\,\trg AP(x)\,\trg BP(y)\,\right>_Q 
 &=& \frac{1}{4}\,\Pi(E;x,y)\,\trg\Bigl[\,AB \no\\
 & & - A\Sigma_zB\Sigma_z -A\Sigma_xB\Sigma_x \no\\
 & & +A\Sigma_yB\Sigma_y-ACB^TC^T \no\\
 & & + A\Sigma_zCB^TC^T\Sigma_z \no\\
 & & +A\Sigma_xCB^TC^T\Sigma_x \no\\
 & & -A\Sigma_yCB^TC^T\Sigma_y \,\Bigr],
\ee
 where $\alpha = +1 (-1)$ for the chiral orthogonal (symplectic) class.
 The presence of the terms including $\Sigma_x$ and $\Sigma_y$ 
 is due to chiral symmetry.
 The terms including the complex conjugation matrix can be interpreted 
 as the cooperon contributions.
 Using these rules, we obtain
\be
 & & \left<\rho(E)\right> 
 = \frac{1}{\Delta}\,\mbox{Re}\,\biggl[\,1+
 2\,\alpha\,\sum_k \tilde{\Pi}(E,k) +\cdots \,\biggr]. \label{1ptos}
\ee
 The first-order contribution appears in contrast with the unitary case.

 If we take the zero mode only, we have
\be 
 \left<\rho(E)\right> 
 &=& \frac{1}{\Delta}\,\mbox{Re}\,\biggl[\, 1 
 +\frac{i\alpha}{2\pi E/\Delta}
% +\frac{5}{16(\pi E/\Delta)^2}
 +\cdots \,\biggr].
\ee
 We see that the $1/E$ term does not contribute to the density of states
 because this term is imaginary.
 On the other hand, this is not the case if one includes  the $k \ne 0$ modes.
 The second term in Eq.(\ref{1ptos}) survives and 
 represents the lowest-order quantum-interference effect.
 We can see the signs in the two classes are opposite,
 which is reminiscent of the weak localization effects of the conductance 
 for the traditional classes. 
 The diffusion mode is essential for this effect to appear. 
%%%%%%%%%%%%%%%%%%%%%%%%%%%%%%%%%%%%%%%%%%%%%%%%%%%%%%%%%%%%%%%%%%%%%%%%%%%%
\subsection{Small-$E$ behavior of the density of states}
 The above results can not describe the small-$E$ behavior
 because we do not treat the zero mode properly.
 For the small energy, the diffusion propagator diverges at the zero momentum, 
 which means the breakdown of the perturbation.
 This defect can be circumvented by 
 considering the zero mode separately.\cite{km,fm} 
 By this improvement, we can discuss the small-$E$ behavior of 
 the level correlation functions that is highly nonperturbative.
 
 The parametrization of $Q$ is as follows;
\be
 Q(x) &=& T_0^{-1}\tilde{Q}(x)T_0.
\ee
 The zero mode $T_0$ is treated exactly \cite{ast} and
 the nonzero modes $\tilde{Q}(x)$ are treated perturbatively as before.

 After the perturbative calculation of the nonzero modes,
 the average density of states is expressed as
\be
 \left<\rho(E)\right> &=& \frac{1}{\Delta}\,\mbox{Re}\int dQ_0\,
 \exp\left[\,\frac{\pi iE}{\Delta}\,\trg P_0^2\,\right] \no\\
 & & \times \biggl[\,1+\frac{1}{2}\,\trg kP_0^2\,\biggr] 
 \biggl[\,1+2\,\Pi^2(0) \no\\
 & & +\frac{2\pi iE}{\Delta}\,\Pi^2(0)\,\trg P_0^2 \no\\
 & & +\frac{4\pi iE}{\Delta}\biggl(\frac{1}{V}\int d^dx\,\Pi^2(x)\biggr)
 \,\trg P_0^2 \no\\
 & & -\frac{2\pi^2E^2}{\Delta^2}
 \biggl(\frac{1}{V}\int d^dx\,\Pi^2(x)\biggr)\,
 \bigl(\,\trg P_0^2\,\bigr)^2 \,\biggr] +\cdots , \no\\ \label{pc2}
\ee
 where $\Pi(x)$ is the diffusion propagator with $E=0$
 and $P_0$ is the supermatrix which parametrizes the zero mode.
 This expansion is valid for $E\ll E_c=D/L^2$ 
 because Eq.(\ref{pc2}) is expanded in powers of $E$.

 The zero mode integration is performed and we obtain
\be
 \frac{\Delta}{\pi}\,\left<\rho(E)\right> 
 &=& \Biggl[\,1+\frac{a_d^{(1)\,2}}{8g^2}
 +\left(\frac{a_d^{(1)\,2}}{8g^2}+\frac{a_d^{(2)}}{4g^2}\right)
 z\frac{\mbox{d}}{\mbox{d} z} \no\\
 & & +\frac{a_d^{(2)}}{8g^2} z^2\frac{\mbox{d}^2}{\mbox{d} z^2}
 +O(1/g^3)\,\Biggr]\,\rho_s(z) \no\\
 &=& \frac{z}{2}\,
 \Bigl[\,J_0^2(z)+J_1^2(z)\,\Bigr]
 + \frac{a_d^{(1)\,2}}{8g^2}\,zJ_0^2(z) \no\\
 & & +\frac{a_d^{(2)}}{8g^2}\,zJ_0(z)\,
 \Bigl[\,J_0(z)-2zJ_1(z)\,\Bigr] \no\\
 & & +O(1/g^3),
\ee
 where $z=\pi E/\Delta,\ g=\pi E_c/\Delta$ 
 and $\rho_s(z)$ is the universal function at the ergodic limit.\cite{vz} 
 Constant $a_d^{(1,2)}$ is the momentum integrations of 
 the diffusion propagator
\be
 a^{(1)}_d &=& \frac{1}{\pi^2}\,
 \sum_{n\ne 0}\,\frac{1}{n_1^2+\cdots +n_d^2}, \\
 a^{(2)}_d &=& \frac{1}{\pi^4}\,
 \sum_{n\ne 0}\,\frac{1}{\bigl(\,n_1^2+\cdots +n_d^2\,\bigr)^2}.
\ee
 These constants depend on the spatial dimension $d$ ;
 $a^{(1)}_1 = 1/6$, 
 $a^{(1)}_2 = \frac{1}{2\pi}\,\ln L/l$, 
 $a^{(1)}_3 \sim L/l$, 
 $a^{(2)}_1 = 1/90 \simeq 0.0111$, 
 $a^{(2)}_2 \simeq 0.0266$, 
 $a^{(2)}_3 \simeq 0.0527$ 
 ($l$ is the mean free path that decides the cutoff scale).
 For some cases, e.g., the renormalizatoin group calculation,
 it is more convenient to include  
 counterterms of higher order in the action (\ref{S})
 instead of using cutoff parameters explicitly.

 For the small $z$ ($z\ll 1 \ll g$), we have
\be
 \frac{\Delta}{\pi}\,\left<\rho(E)\right>  
 &\sim& \frac{1}{2}\,\left[\,1+\frac{a_d^{(1)\,2}}{4\,g^2}
 +\frac{a_d^{(2)}}{4\,g^2}\,\right]\,z. 
 \label{smallz}
\ee
 For $1 \ll z \ll g$,
\be
 \Delta\left<\rho(E)\right>  
 &\sim& 1+\frac{a_d^{(1)\,2}}{8\,g^2}-\frac{a_d^{(2)}}{4\,g^2}.
 \label{largez}
\ee
 Eq.(\ref{largez}) can also be obtained from Eq.(\ref{1pt}).

 We can perform the same calculations for the other classes.
 In this case, there remains a $1/g$ term and we get
\be
 \frac{\Delta}{\pi}\,\left<\rho(E)\right> 
 &=& \biggl[\,1+\alpha\,\frac{a_d^{(1)}}{4g}
 \Bigl(1+z\frac{\mbox{d}}{\mbox{d}z}\Bigr)
 +O(1/g^2) \,\biggr]\,\rho_s(z). \no\\
\ee
 It is remarkable that for all three classes the deviations can 
 be expressed with the universal function $\rho_s(z)$ and 
 its derivative. 
 Whether this also holds for higher-order terms seems an interesting 
 problem.
 The universal functions $\rho_s(z)$ 
 for the orthogonal and symplectic cases are 
 already derived by the orthogonal polynomial method.\cite{goe,gse} 

 For the small $z$, we have
\be
 \frac{\Delta}{\pi}\,\left<\rho(E)\right> &\sim& 
 \frac{1}{2}\,\left(\,1+\frac{a_d^{(1)}}{2\,g}\,\right),\ \mbox{orthogonal},
 \label{smallzo} \\
 \frac{\Delta}{\pi}\,\left<\rho(E)\right> &\sim& 
 \frac{1}{3}\,\left(\,1-\frac{2a_d^{(1)}}{g}\,\right)\,z^3,\ \mbox{symplectic}. 
 \label{smallzs}
\ee
 For $1 \ll z \ll g$,
\be
 \Delta\left<\rho(E)\right> &\sim& 
 1+\alpha\,\frac{a_d^{(1)}}{4\,g},
\ee
 which is again consistent with Eq.(\ref{1ptos}).

 Equations (\ref{smallz}) and (\ref{smallzo}), and (\ref{smallzs})
 mean that the resulting correction 
 does not change the power behavior of the density of states, 
 but renormalizes the corresponding prefactor.
 Similar expressions exist for the two-point level-correlation function 
 for the traditional classes.\cite{km}  
 We obtain the results for three traditional classes
 that are interpreted as a weakening of the level repulsion.
 The results for the chiral unitary and orthogonal classes could be  
 interpreted similarly as a weakening of the repulsion between 
 a pair of levels with the energy $\pm E$.
 The result for the symplectic case does not accept this 
 trivial interpretation and is interesting.

 Gade analyzed a similar model in the thermodynamic limit by 
 the renormalization group method for the chiral unitary class and
 got the divergence at the band center \cite{gade} for $d=2$.
 Our analysis implies Gade's result.
 It is interesting to perform the renormalization group analysis 
 for the chiral symplectic class
 since our result suggests the vanishing at the band center.

 We note the dimensional dependence of the conductance $g$.
 For $\Delta\sim L^{-d}$ and $E_c\sim L^{-2}$, we have $g\sim L^{d-2}$.
 Therefore, our $1/g$ corrections become irrelevant for $d>2$
 in the thermodynamic limit.

%%%%%%%%%%%%%%%%%%%%%%%%%%%%%%%%%%%%%%%%%%%%%%%%%%%%%%%%%%%%%%%%%%%%%
\subsection{2pt. level correlation function}
 We calculate the large-E behavior of the two-point level-correlation function.
 In this case, the saddle-point manifold is parametrized as
\be
 Q(x) = -i\Sigma_z(1+P)(1-P)^{-1}, \\
 P = \pmatrix{
 0 & it(x) \cr -it(x) & 0 \cr},\ 
 t(x) = \pmatrix{
 t_1(x) & t_{12}(x) \cr
 t_{21}(x) & t_2(x) \cr}.
\ee
 Here, $2\times 2$ supermatrices $t_1$ and $t_2$ are due to 
 the chiral structure of the model and
 $t_{12}$ and $t_{21}$ are due to the calculation of 
 the two-point function.\cite{th,ast} 

 For the unitary class, the leading-order contribution of the connected part 
 of the level correlation function 
 comes from the following contraction:\cite{th} 
\be
 \Bigl<\,\trg kt_{12}t_{21}(x)\,
 \trg kt_{21}t_{12}(y)\,\Bigr>_Q 
 = 4\,\Pi^2(E;x,y).
\ee
 We have
\be
 & & \Delta^2 \Bigl[\,\left<\,\rho(E_1)\rho(E_2)\,\right>
 -\left<\,\rho(E_1)\,\right>\left<\,\rho(E_2)\,\right>\,\Bigr] \no\\
 &=& \sum_k \,
 \Bigl[\,\tilde{\Pi}^2(E,k)+\tilde{\Pi}^2(-E,k) \no\\
 & & +\tilde{\Pi}^2(\omega/2,k)+\tilde{\Pi}^2(-\omega/2,k)\,\Bigr]
 +\cdots . \label{2pt}
\ee
 We see that the last two terms already appear in  
 the ordinary unitary class.\cite{as} 
 The first two terms are characteristic to the chiral class.
 Due to this contribution,
 the number variance of the chiral unitary class is
 two times of that of the unitary class in this calculation.\cite{th} 

 If we take the zero mode only,
\be 
 & & \Delta^2\Bigl[\,\left<\,\rho(E_1)\rho(E_2)\,\right>
 -\left<\,\rho(E_1)\,\right>\left<\,\rho(E_2)\,\right>\,\Bigr] \no\\
 &=& -\frac{1}{2}\,\biggl[\,\frac{1}{4(\pi E/\Delta)^2}
 +\frac{1}{(\pi \omega/\Delta)^2}\,\biggr], \label{2pt0}
\ee
 which is the asymptotics of the exact result at the ergodic limit.\cite{vz} 
 
 For the chiral orthogonal and symplectic classes, this function becomes 
 two times larger than that of the chiral unitary class.
 This is due to the cooperon contribution.
 This feature is almost the same as the traditional classes. 
%%%%%%%%%%%%%%%%%%%%%%%%%%%%%%%%%%%%%%%%%%%%%%%%%%%%%%%%%%%%%%%%%%%%%%%%
%%%%%%%%%%%%%%%%%%%%%%%%%%%%%%%%%%%%%%%%%%%%%%%%%%%%%%%%%%%%%%%%%%%%%%%%
\section{Diagrammatic perturbation theory}
 In this section, we reexamine the results obtained above 
 with the diagrammatic method.
 We are especially interested in the kind of diagrams that make 
 chiral disordered systems different from traditional ones.
 This will give us some insights into the kind of quantities 
 that are characteristic to the chiral classes.
 We work with the ergodic limit for simplicity.
 The generalization to the diffusive regime can be done 
 along the same line as the traditional classes. 
 Similar calculations have been done for 
 systems with particle-hole symmetry.\cite{az} 

 We can summarize the Gaussian correlation law of the chiral symmetric 
 Hamiltonian for each classes as
\be
 & & \left<H_{ij}\,H_{kl}\right>
 = \frac{v^2}{2}\ \Bigl[\,\delta_{il}\,\delta_{kj}
 -(\Sigma_z)_{il}\,(\Sigma_z)_{kj}\,\Bigr],\ \mbox{unitary},\\
 & & \left<H_{ij}\,H_{kl}\right> 
 = \frac{v^2}{4}\ \Bigl[\,\delta_{il}\,\delta_{kj}
 +\delta_{ik}\,\delta_{lj} \no\\
 & & -(\Sigma_z)_{il}\,(\Sigma_z)_{kj}
 -(\Sigma_z)_{ik}\,(\Sigma_z)_{lj}\,\Bigr],\ \mbox{orthogonal},\\
 & & \left<H_{ij}\,H_{kl}\right> 
 = \frac{v^2}{4}\ \Bigl[\,\delta_{il}\,\delta_{kj}
 +(\sigma_y)_{ik}\,(\sigma_y)_{lj} \no\\
 & & -(\Sigma_z)_{il}\,(\Sigma_z)_{kj}
 -(\Sigma_z\,\sigma_y)_{ik}\,(\sigma_y\,\Sigma_z)_{lj}\,\Bigr],\ 
 \mbox{symplectic}.\label{gauss-s}
\ee
 The effects of chiral symmetry are characterized 
 by the presence of the terms including $\Sigma_z$.

 First, summing the rainbow-type diagrams, we can get the familiar form
 of the Green function as
\be
 \left<G^{(s)}(E)\right> &=& \frac{\pi^2E}{8N^2\Delta^2}
 -is\,\frac{\pi}{2N\Delta}\sqrt{1-\left(\frac{\pi E}{4N\Delta}\right)^2},
 \label{1ptg}
\ee
 where $s=+1 (-1)$ for the retarded (advanced) function
 and $2N$ is the size of the Hamiltonian.

 We know that the quantum-interference effects are expressed 
 by the diffuson and cooperon diagrams.
 The diffuson diagram is calculated as
\be
 & & \Pi_{ij,kl}^{(s,s')}(E_1,E_2) \no\\
 &=& \frac{2N\Delta^2}{\pi^2}\Biggl[\,
 \frac{\delta_{il}\,\delta_{kj}}
 {1-\left(\frac{2N\Delta}{\pi}\right)^2
 \left<G^{(s)}(E_1)\right>\left<G^{(s')}(E_2)\right>}\no\\
 & & -\frac{(\Sigma_z)_{il}\,(\Sigma_z)_{kj}}
 {1+\left(\frac{2N\Delta}{\pi}\right)^2
 \left<G^{(s)}(E_1)\right>\left<G^{(s')}(E_2)\right>}
 \,\Biggr].
\ee
 Using the Green function (\ref{1ptg}), we have
\be
 & & \Pi_{ij,kl}^{(+,+)}(E_1,E_2) \no\\ 
 &=& \left(\frac{2N\Delta}{\pi}\right)^2\, 
 \biggl[\,\frac{1}{4N}\,\delta_{il}\delta_{kj} 
 -i \frac{(\Sigma_z)_{il}\,(\Sigma_z)_{kj}}{\pi E/\Delta}\,\biggr], \\
 & & \Pi_{ij,kl}^{(-,-)}(E_1,E_2) \no\\ 
 &=& \left(\frac{2N\Delta}{\pi}\right)^2\, 
 \biggl[\,\frac{1}{4N}\,\delta_{il}\delta_{kj} 
 +i \frac{(\Sigma_z)_{il}\,(\Sigma_z)_{kj}}{\pi E/\Delta}\,\biggr], \\
 & & \Pi_{ij,kl}^{(+,-)}(E_1,E_2) \no\\
 &=& \left(\frac{2N\Delta}{\pi}\right)^2\, 
 \biggl[\,i \frac{\delta_{il}\delta_{kj}}{\pi \omega/2\Delta}
 -\frac{1}{4N}\,(\Sigma_z)_{il}\,(\Sigma_z)_{kj}\,\biggr], \\
 & & \Pi_{ij,kl}^{(-,+)}(E_1,E_2) \no\\
 &=& \left(\frac{2N\Delta}{\pi}\right)^2\, 
 \biggl[\,-i \frac{\delta_{il}\delta_{kj}}{\pi \omega/2\Delta}
 -\frac{1}{4N}\,(\Sigma_z)_{il}\,(\Sigma_z)_{kj}\,\biggr],
\ee
 where $E=(E_1+E_2)/2$ and $\omega=E_1-E_2$.
 There are singular contributions ($1/E$ or $1/\omega$ terms) 
 and non-singular contributions.
 The remarkable property is that singular $1/E$ terms appear even 
 in the retarded-retarded (advanced-advanced) channel.

 For the chiral orthogonal class, we have the cooperon contribution
\be
 & & \bar{\Pi}_{ij,kl}^{(s,s')}(E_1,E_2) \no\\
 &=& \frac{2N\Delta^2}{\pi^2}\Biggl[\,
 \frac{\delta_{ik}\,\delta_{lj}}
 {1-\left(\frac{2N\Delta}{\pi}\right)^2
 \left<G^{(s)}(E_1)\right>\left<G^{(s')}(E_2)\right>}\no\\
 & & \qquad\qquad -\frac{(\Sigma_z)_{ik}\,(\Sigma_z)_{lj}}
 {1+\left(\frac{2N\Delta}{\pi}\right)^2
 \left<G^{(s)}(E_1)\right>\left<G^{(s')}(E_2)\right>}
 \,\Biggr].
\ee
 This is the same as the diffuson contribution 
 except the indices $k \leftrightarrow l$.
 For the chiral symplectic class, 
 the Pauli matrix $\sigma_y$ is inserted in 
 the cooperon contribution as Eq.(\ref{gauss-s}).
%%%%%%%%%%%%%%%%%%%%%%%%%%%%%%%%%%%%%%%%%%%%%%%%%%%%%%%%%%%%%%%%%%%%%%%%
\subsection{Density of states}
\begin{figure}
\begin{center}
\leavevmode
\epsfxsize=6cm \epsfbox{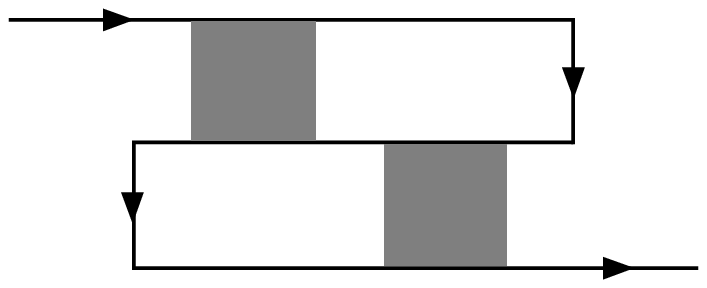}
\end{center}
\caption{Diagrams contributing to the Green function 
 for the chiral unitary class.
 The dark-shaded regions represent a diffuson-type ladder.}
\label{fig:1ptu}
\end{figure}

 For the chiral unitary class, only the diffuson-type ladder appears.
 The diagram for the lowest order correction of the Green 
 function is depicted in Fig.\ref{fig:1ptu}.
 The diagrams of this order consist of two singular-diffuson diagrams  
 and the same ones with nonsingular corrections. 
 (In Fig.\ref{fig:1ptu}, we do not write the nonsingular contributions 
 that consist of nine diagrams.)
 We get the Green function 
\be
 \Bigl<\,\tr G^{(R)}(E)\,\Bigr> 
 = -i\,\frac{\pi}{\Delta}\,\biggl[\,1
 +\frac{1}{8(\pi E/\Delta)^2}+\cdots \,\biggr],
\ee
 which coincides with the supersymmetric calculation.
%---------------- fig1 ----------------

\begin{figure}
\begin{center}
\leavevmode
\epsfxsize=8cm \epsfbox{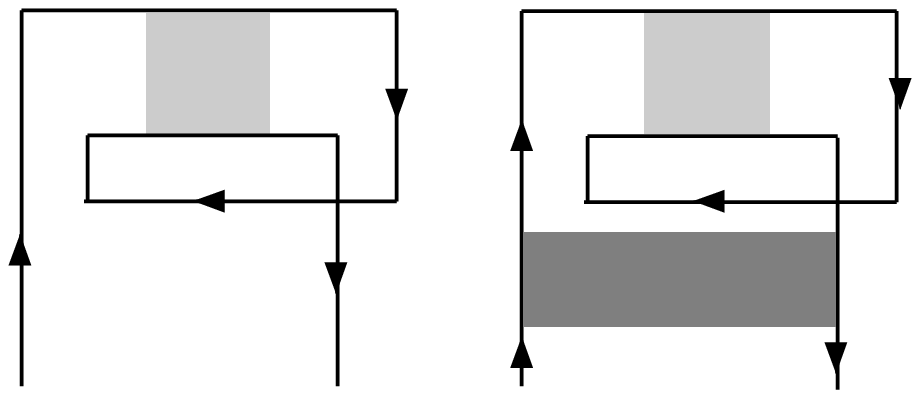}
\end{center}
\caption{Diagrams contributing to the Green function 
 for the chiral orthogonal and symplectic classes.
 The light-shaded regions represent a cooperon-type ladder.}
\label{fig:1ptos}
\end{figure}

 For the chiral orthogonal and symplectic classes, we can anticipate 
 the $1/E$ term contribution.
 This is indeed the case and we can express the $1/E$ term contribution
 as Fig.\ref{fig:1ptos}.
 We note that the singular contribution comes from the cooperon-type ladder
 and the diffusion ladder contribution is the nonsingular one.
 We obtain the Green function as
\be
 \Bigl<\tr G^{(R)}(E)\Bigr> 
 &=& -i\,\frac{\pi}{\Delta}\,\biggl[\,1
 +\frac{i\alpha}{2\pi E/\Delta} \no\\
 & & +\frac{5}{16(\pi E/\Delta)^2}+\cdots \,\biggr].
\ee
 We confirm that the first three terms
 coincide again with the supersymmetric calculation.
%---------------- fig2 ----------------
%%%%%%%%%%%%%%%%%%%%%%%%%%%%%%%%%%%%%%%%%%%%%%%%%%%%%%%%%%%%%%%%%%%%%%%%
\subsection{Two-point level-correlation function}
 The leading order contributions 
 to the twp-point level-correlation function can be expressed with 
 the same diagram as the traditional classes 
 (two diffusons and two cooperons).\cite{as} 
 For the chiral unitary case, we take the diffuson-type diagram only.
 Calculating these diagrams, we can derive Eq.(\ref{2pt0}).
%%%%%%%%%%%%%%%%%%%%%%%%%%%%%%%%%%%%%%%%%%%%%%%%%%%%%%%%%%%%%%%%%%%%%%%%
%%%%%%%%%%%%%%%%%%%%%%%%%%%%%%%%%%%%%%%%%%%%%%%%%%%%%%%%%%%%%%%%%%%%%%%%
\section{Conclusions}
 In this paper, we have formulated the impurity perturbative expansion
 for chiral symmetric disordered systems.
 For three chiral classes,
 we calculated the large-$E$ and small-$E$ behavior 
 of the density of state and 
 the large-$E$ behavior of the two-point level-correlation function.
 The deviations from the universal function are obtained in which   
 we find the effects similar to the weak localization effects 
 for the traditional classes. 
 We calculate these deviations by considering the nonzero-$k$ modes 
 explicitly.
 Alternatively, one can, in principle, integrate out 
 the nonzero modes from the start
 and get the effective action for the zero mode.
 Then, working with this action should give the same results.

 For the behavior at the large energy, 
 we can interpret the results diagrammatically. 
 We see that the diffuson and cooperon modes 
 generate the nontrivial contribution
 to the level correlation functions.
 We have the diffuson and cooperon contributions in the advanced-advanced 
 (retarded-retarded) channel in addition to the usual ones.
 
 For the chiral unitary and orthogonal classes 
 we get the small-$E$ behavior that could be  
 interpreted as a weakening of the repulsion between 
 a pair of levels with the energy $\pm E$.
 For the chiral symplectic case, however, we get 
 a result with opposite tendency whose intuitive explanation 
 remains an open problem.
 These results imply the divergence of the density of states 
 at the band center for the chiral unitary and orthogonal classes and 
 the vanishing for the chiral symplectic class
 in the thermodynamic limit.
 For the chiral unitary class, 
 this divergence is observed by the renormalization
 group calculation.\cite{gade} 
 The calculations for the other classes are future problems.
 
 If once we derive the nonlinear sigma model and 
 formulate the perturbation theory,
 it is easy to use this calculation to several applications
 such as the study of the wave function statistics and 
 the renormalization group equations.
 These are currently under intensive study.
%%%%%%%%%%%%%%%%%%%%%%%%%%%%%%%%%%%%%%%%%%%%%%%%%%%%%%%%%%%%%%%%%%%%%%%%
\section*{Acknowledgments}
 We are grateful to N. Taniguchi for valuable discussions
 and S. Higuchi for a careful reading of the manuscript.
 K.T is supported financially by 
 the Japan Society for the Promotion of Science for Young Scientists.
%%%%%%%%%%%%%%%%%%%%%%%%%%%%%%%%%%%%%%%%%%%%%%%%%%%%%%%%%%%%%%%%%%%%%%%%
%%%%%%%%%%%%%%%%%%%%%%%%%%%%%%%%%%%%%%%%%%%%%%%%%%%%%%%%%%%%%%%%%%%%%%%%

\widetext
\end{document}